\documentclass[aps,twocolumn,amsmath,amssymb,showpacs,prl,superscriptaddress,unsortedaddress]{revtex4}
\usepackage{epsf}
\usepackage{graphicx}
\usepackage{sidecap}
\usepackage{amsmath}
\usepackage{float}
 \usepackage{gensymb}
 \usepackage{hyperref}

\usepackage{color}
 
\def \beq {\begin{equation}}
\def \eeq {\end{equation}}
\begin{document}
 
\title{{Temperature Dependent Electronic Structure in a Higher Order Topological Insulator Candidate EuIn$_{2}$As$_{2}$}}
 \author{Sabin Regmi} \affiliation {Department of Physics, University of Central Florida, Orlando, Florida 32816, USA}
 \author{Md~Mofazzel~Hosen}\affiliation {Department of Physics, University of Central Florida, Orlando, Florida 32816, USA}
 \author{Barun~Ghosh}\affiliation{Department of Physics, Indian Institute of Technology Kanpur, Kanpur 208016, India}
 \author{Bahadur~Singh}\affiliation{Department of Physics, Northeastern University, Boston, Massachusetts 02115, USA} \affiliation{SZU-NUS Collaborative Center and International Collaborative Laboratory of 2D Materials for Optoelectronic Science \& Technology, Engineering Technology Research Center for 2D Materials Information Functional Devices and Systems of Guangdong Province, Institute of Microscale Optoelectronics, Shenzhen University, Shenzhen 518060, China}
\author{Gyanendra~Dhakal}\affiliation {Department of Physics, University of Central Florida, Orlando, Florida 32816, USA}
\author{Christopher~Sims}\affiliation {Department of Physics, University of Central Florida, Orlando, Florida 32816, USA}
\author{Baokai~Wang}\affiliation{Department of Physics, Northeastern University, Boston, Massachusetts 02115, USA}
\author{Firoza~Kabir}\affiliation {Department of Physics, University of Central Florida, Orlando, Florida 32816, USA}
\author{Klauss~Dimitri}\affiliation {Department of Physics, University of Central Florida, Orlando, Florida 32816, USA}  
\author{Yangyang Liu}\affiliation {Department of Physics, University of Central Florida, Orlando, Florida 32816, USA}  
\author{Amit~Agarwal}\affiliation{Department of Physics, Indian Institute of Technology Kanpur, Kanpur 208016, India}
\author{Hsin~Lin}\affiliation{Institute of Physics, Academia Sinica, Taipei 11529, Taiwan}
 \author{Dariusz~Kaczorowski}\affiliation {Institute of Low Temperature and Structure Research, Polish Academy of Sciences, 50-950 Wroclaw, Poland}
 \author{Arun~Bansil}\affiliation{SZU-NUS Collaborative Center and International Collaborative Laboratory of 2D Materials for Optoelectronic Science \& Technology, Engineering Technology Research Center for 2D Materials Information Functional Devices and Systems of Guangdong Province, Institute of Microscale Optoelectronics, Shenzhen University, Shenzhen 518060, China}
 \author{Madhab~Neupane}
\affiliation {Department of Physics, University of Central Florida, Orlando, Florida 32816, USA}
 
\date{6 October, 2019}

\begin{abstract}
\noindent

The higher order topological insulator (HOTI) has enticed enormous research interests owing to its novelty in supporting gapless states along the hinges of the crystal. Despite several theoretical predictions, enough experimental confirmation of HOTI state in crystalline solids is still lacking. It has been well known that interplay between topology and magnetism can give rise to various magnetic topological states including HOTI and Axion insulator states. Here using the high-resolution angle-resolved photoemission spectroscopy (ARPES) combined with the first-principles calculations, we report a systematic study on the electronic band topology across the magnetic phase transition in EuIn$_2$As$_2$ which possesses an antiferromagnetic ground state below 16 K. Antiferromagnetic EuIn$_2$As$_2$ has been predicted to host both the Axion insulator and HOTI phase. Our experimental results show the clear signature of the evolution of the topological state across the magnetic transition. Our study thus especially suited to understand the interaction of higher-order topology with magnetism in materials.


\end{abstract}

\date{\today}
\maketitle
\noindent \textit{Introduction.} - The discovery of topological insulator \cite{hk,zhangrev,hasan,bansil} began the era of loads of theoretical and experimental works on various  topological quantum materials \cite{zrsis, taas, hosen, klauss, cd3as2}. The topological insulators are driven by the conventional bulk-boundary correspondence, meaning these materials possess insulating bulk, but there are gapless conducting states on the surface. These surface states disperse linearly around a point - called the Dirac point, are spin polarized, and the presence of time-reversal symmetry protects them from back-scattering and localization in the presence of weak perturbations, thereby making these insulating quantum materials suitable for potential application in low-power energy efficient quantum electronic devices \cite{wolf, xlqi, linder}. Recently, new kind of topological insulating materials are on the scene, which do not exhibit the usual bulk-boundary correspondence, instead possess bulk-surface-hinge correspondence in which  along with the insulating bulk, the surface  is also gapped with finite mass term, and the material hosts gapless topological states along the hinges. These materials are called higher order topological insulators (HOTIs) \cite{huges, benalcazar, neupert}. The surfaces adjoining the hinges are required to be in different topological classes \cite{langevin}.

\noindent The higher order topological insulating phase is perceived to be protected by different symmetries like mirror symmetry \cite{neupert}, rotation symmetry \cite{song, benalcazar}, reflection symmetry \cite{langevin}, inversion symmetry \cite{khalaf}, etc. Such phases have already been realized in electronic circuits \cite{imhof}, photonic systems \cite{peterson}, and phononic systems \cite{serra}. Recent work has reported higher order topology to exist in 3D crystalline Bismuth via combination of first-principles calculations, scanning-tunneling spectroscopy (STM), and Josephson Interferometry \cite{bismuth}. In addition to that, HOTI phase has been proposed to be realized in  EuIn$_2$As$_2$ \cite{dai}, and Sm-doped Bi$_2$Se$_3$ \cite{dai2}, and several other materials \cite{neupert, ezawa, ezawa1, xue}. Till date, no momentum resolved spectroscopic measurements have been performed on HOTI candidate materials.

 \begin{figure*}[htbp!]
	\centering
	\includegraphics[width=16 cm]{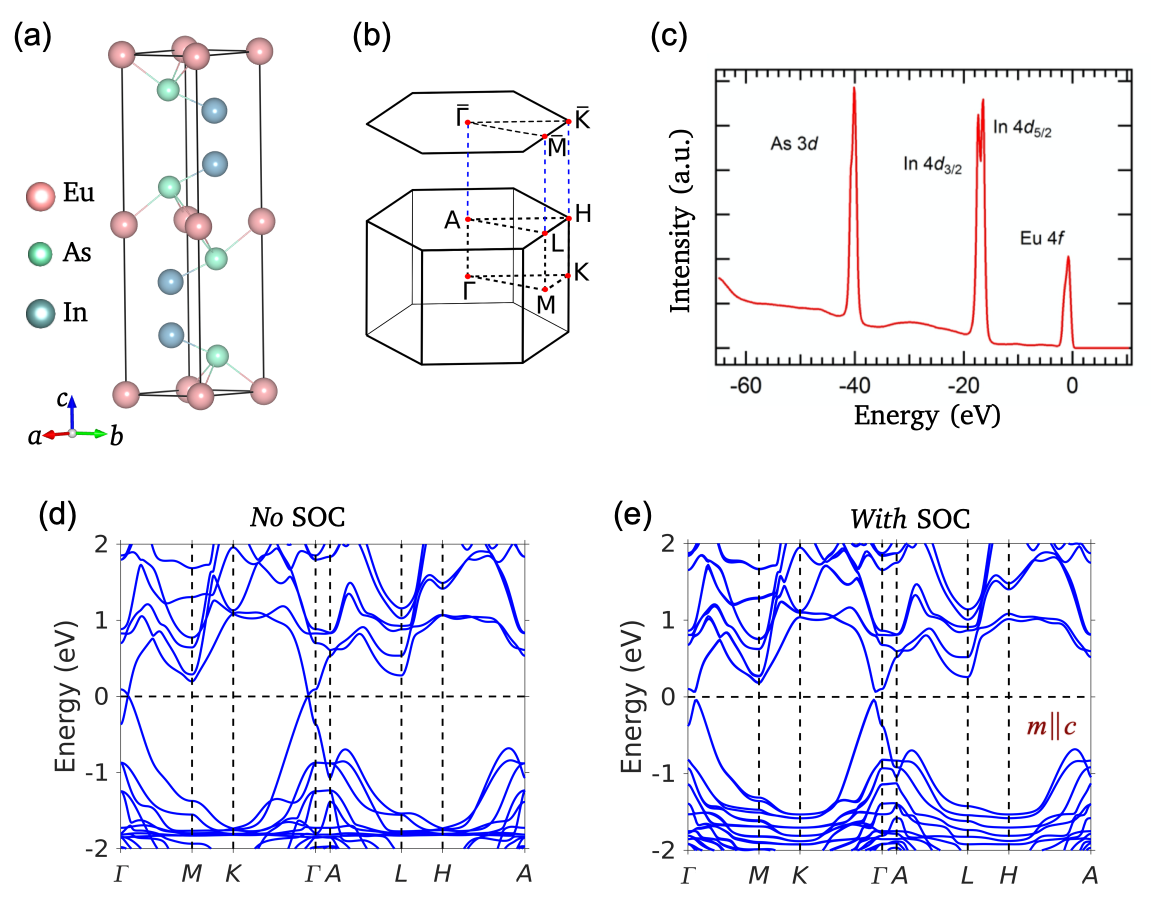}
	\caption{Crystal structure and sample characterization of  EuIn$_2$As$_2$.
(a) Crystal structure of EuIn$_2$As$_2$. Light red, light green, and gray balls identify Eu, As, and In atoms, respectively.  (b) The associated bulk and (001) surface Brillouin zones. The high symmetry points are marked. (c) Core-level photoemission spectrum with characteristic peaks of Eu 4$f$, In 4$d$, and As 3$d$ orbitals. (d) Calculated bulk bands along the bulk high-symmetry directions without spin-orbit coupling (SOC). (e) Calculated bulk bands taking into account the SOC effect.  }
\end{figure*} 

\noindent Topology incorporated with magnetism can give rise to a variety of novel quantum phenomena. The introduction of spontaneous magnetization in a topological insulator opens  a gap at the Dirac point, and if the Fermi level is tuned within this gap, phenomenon such as quantum Hall effect occurs \cite{tokura}. Furthermore, the interplay between topology and magnetism may lead to novel phases such as axion insulator \cite{axion, zhang}, antiferromagnetic topological insulator \cite{ otrokov,chen, ding,  hao, hu}, and other magnetic topological quantum states \cite{gdsbte, ilya, liu, gui}.

\noindent In this article, we report the electronic structure of EuIn$_2$As$_2$ which possesses magnetism and is a potential ground to host topological states based on previous studies.  We use high-resolution ARPES and parallel first-principles calculations to study the detailed electronic structure of EuIn$_2$As$_2$.  Our theoretical calculations predict this material to be an axion  insulator  in both the  AFM-B phase (in which magnetic moments are \textit{in-plane}) and  the  AFM-C phase (in which magnetic moments are \textit{out-of-plane}). Our experimental data clearly show the evolution of the band structure in the vicinity of the Fermi level across the magnetic transition temperature. Our study would pave a pathway to understand the interplay between magnetism and  topology in this HOTI candidate material.

\begin{figure*} [htbp!]
	\includegraphics[width=16 cm]{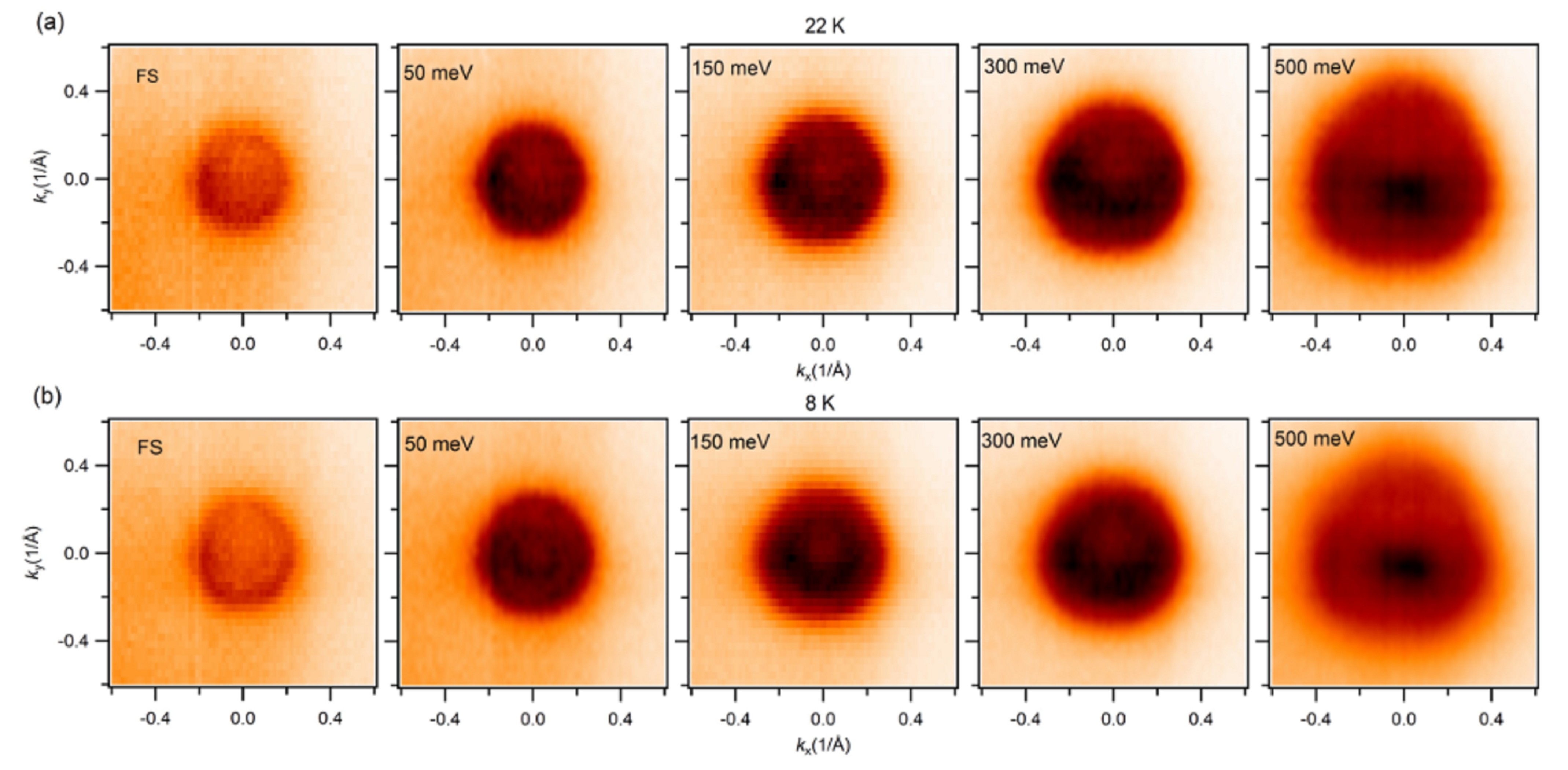}
	\caption{ Fermi surface  map and constant energy contours.
(a) Fermi surface map (leftmost) and constant energy contours measured at a temperature of 22 K. (b) Fermi surface map (leftmost) and constant energy contours measured at a temperature of 8 K. Binding energies are noted on each plots. All the data were collected at the ALS beamline 10.0.1 using photon energy of 60 eV.}
\end{figure*}
 
\noindent \textit{Methods and Experimental Details} -- The single crystals of  EuIn$_2$As$_2$ were grown by using flux method which is described elsewhere \cite{zrsis}. Their chemical composition and crystal structure were examined by energy-dispersive X-ray spectroscopy and single-crystal X-ray diffraction (XRD) respectively. The ARPES measurements were performed by using synchrotron source at the Advanced Light Source (ALS), Berkeley at Beamline 10.0.1  equipped with a high efficiency R4000 electron analyzer. The energy resolution was set better than 20 meV and the angular resolution was set better than 0.2$\degree$.
The samples were cleaved in situ and measured between 8 K - 50 K in a vacuum better than 10$^{-10}$ torr. The crystals were very stable for the typical measurement period of
20 hours. In order to reveal the nature of the states observed
in EuIn$_2$As$_2$, the ARPES data were compared with
the calculated band dispersion projected onto a 2D Brillouin
zone (BZ). We performed the first-principles calculations within the framework of density functional theory (DFT) using the projector-augmented-wave (PAW) pseudopotentials as implemented in the VASP \cite{kohn, kresse, kresse2}. The generalized gradient approximation(GGA) was used to incorporate exchange-correlation effects \cite{perdew}. The on-site Coulomb interaction was added for Eu $f$-electrons within the GGA+U scheme with U$_{eff}$ = 5 eV \cite{dudarev}. An energy cutoff of 400 eV was used for the plane wave basis set and a $\Gamma$-centred 12$\times$12$\times$6 k-mesh was employed for the bulk Brillouin zone integration. The topological state properties were calculated by employing the first-principles tight-binding Hamiltonian which is generated using VASP2WANNIER90 interface \cite{marzari}. The surface energy spectrum was obtained within the iterative Green's function scheme via employing the WannierTools package \cite{wu}.

\noindent \textit{Results and Discussion} - EuIn$_2$As$_2$ crystallizes in a hexagonal lattice, as shown in Fig. 1(a), with the space group of P6$_3/mmc$ (\# 194). It possesses antiferomagnetic ground state below the antiferromagnetic transition temperature T$_N$ = 16 K \cite{goforth, singh}.  In Fig. 1(b), we show the three-dimensional (3D) bulk Brillouin zone (BZ) and its projection onto the two-dimensional (2D) [001] surface. Figure 1(c) represents the core level photoemission spectrum which clearly manifests the characteristic peaks coming from Eu 4$f$, In 4$d$, and As 3$d$ orbitals.
The calculated band structure of antiferromagnetic EuIn$_2$As$_2$ is presented in Fig. 1(d)-(e). In the absence of spin-orbit coupling (SOC), the valence and conduction bands dip into each other establishing a clear band inversion with an inverted bandgap of ~0.46 eV at the $\Gamma$ point [see Fig. 1(d)]. The orbital resolved band structure further shows that the band inversion happens between In \textit{s} and As \textit{p} states [see supplementary information]. While the band structure resembles band crossings feature around the $\Gamma$-K and $\Gamma$-M directions, we find that the bands along the $\Gamma$-M direction are gapped whereas they are gapless along the $\Gamma$-K direction and symmetry protected \cite{dai}. After including SOC, all the band crossings are gapped, separating valence and conductions bands [Fig. (d), also see supplementary information]. Notably, we consider both antiferromagnetic configurations,  AFM-B  and AFM-C. We find that both the AFM-B and AFM-C states are almost degenerate with the calculated energy difference of less than 1 meV in agreement with earlier theory results \cite{dai} [see supplementary information].

\begin{figure*}[htbp!]
	\centering
	\includegraphics[width=16 cm]{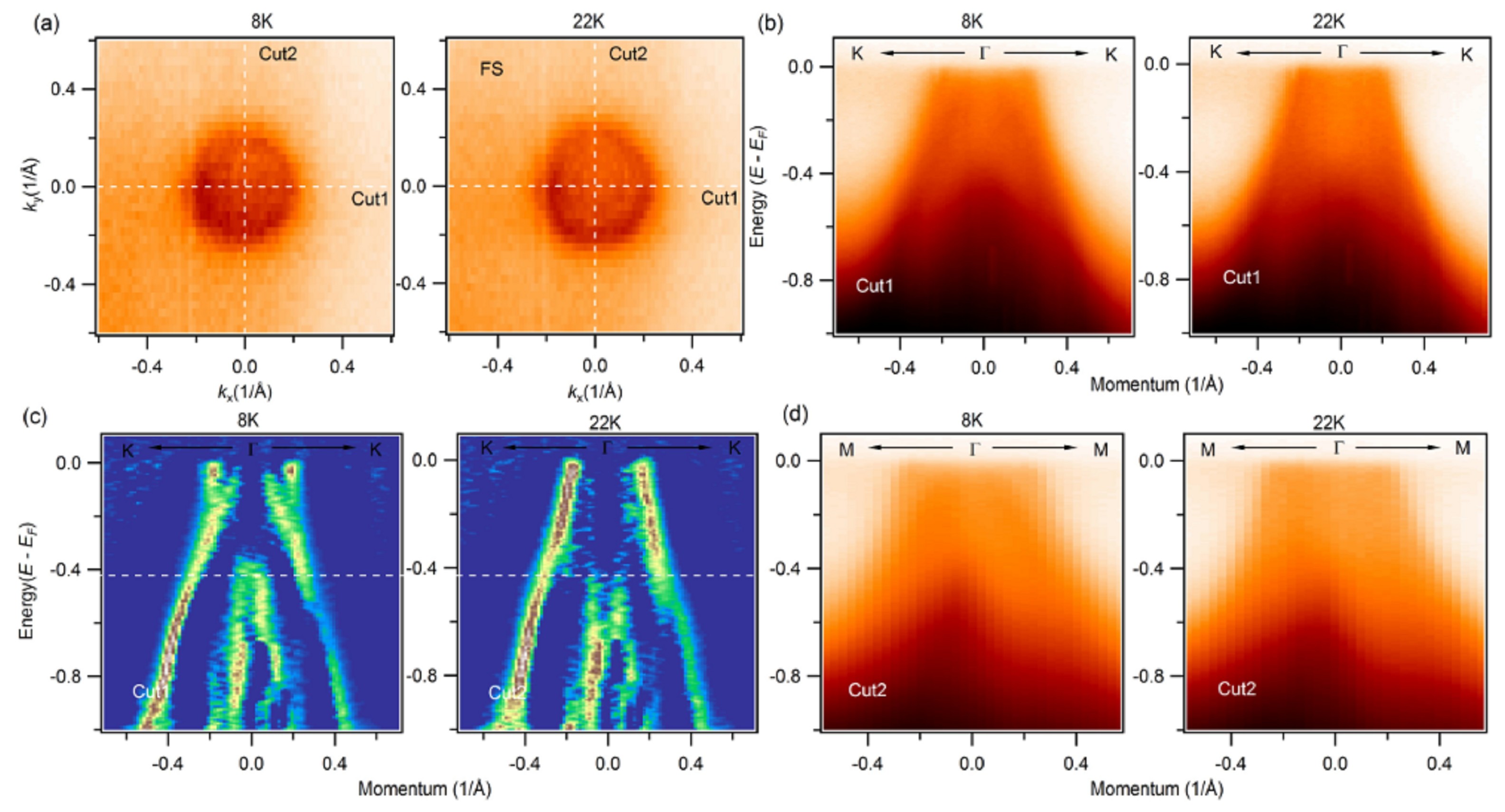}
	\caption{Electronic band structure along different momentum directions. (a) Fermi surface maps with the white dashed lines showing the different cut directions in which the dispersion maps were taken. (b) Dispersion maps along cut 1 (K-$\Gamma$-K) directions in Fig. 2(a). (c) Second derivative plot of dispersion maps in Fig. 2(b). (d) Dispersion maps along cut 2 (M-$\Gamma$-M) directions in Fig. 2(a). The temperatures are noted on each plots. All the data were taken at the ALS beamline 10.0.1 using photon energy of 60 eV.}
\end{figure*}

\noindent Since both AFM-B and AFM-C are gapped with a band inversion, we determine the topological state for both the configurations by calculating the parity-based $\mathbb{Z}_4$ topological invariant \cite{dai, turner, ono, watanabe} which is defined as:

\begin{equation}
\mathbb{Z}_4 = \sum_{\alpha=1}^{8} \sum_{n=1}^{N_{occ}} \frac{1 + \zeta_n(\Gamma_i)}{2}  mod 4
\end{equation}

Here, $\zeta_n$ ($\Gamma_i$) is the parity at the i$^{th}$ time-reversal invariant momenta (TRIM) point $\Gamma_i$ for the n$^{th}$ band, and N$_{occ}$ is the number of occupied bands. We find $\mathbb{Z}_4$ = 2 for both the AFM-B and AFM-C configurations which indicate that EuIn$_2$As$_2$ is an Axion insulator.

\begin{figure*}[htbp!]
	\centering
	\includegraphics[width=16 cm]{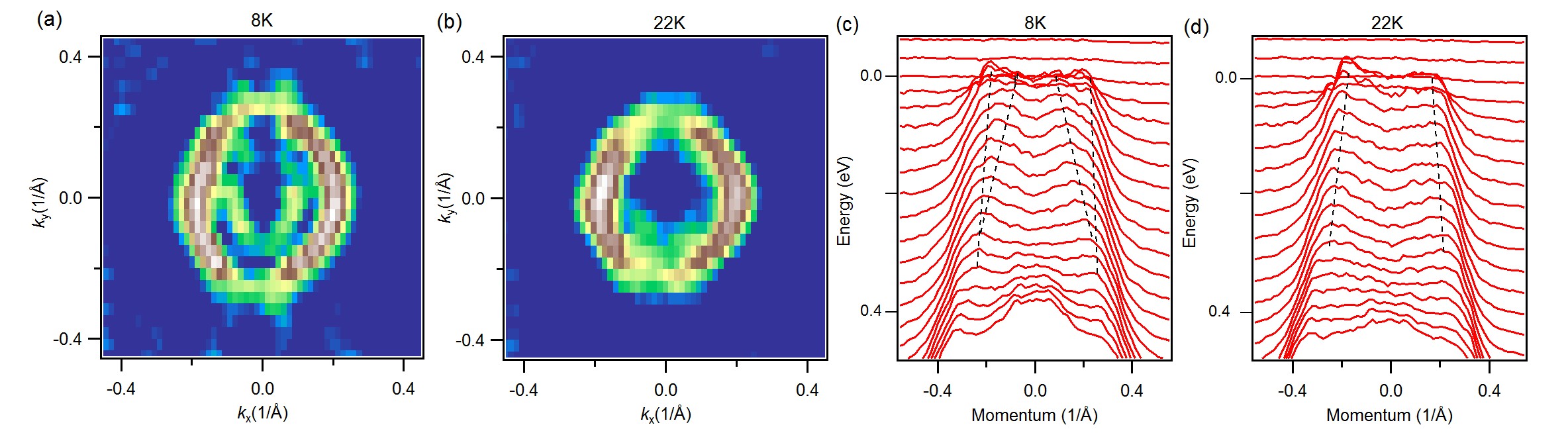}
	\caption{Temperature dependent electronic structure. (a) Second derivative plots of the constant energy contours at the binding energy of 50 meV at 8 K (left) and 22 K (right). (b) MDCs in the vicinity of Fermi level along the K-$\Gamma$-K direction in AFM phase (left) and paramagnetic phase (right).}
\end{figure*}

\noindent Next, we  discuss the ARPES data starting from the Fermi surface and constant energy contour plots taken at different binding energies in both paramagnetic and AFM phases. The leftmost plot in Fig. 2(a) shows the Fermi surface map taken at a temperature of 22 K followed by the energy contours at the binding energies of 50 meV, 150 meV, 300 meV, and 500 meV. Figure 2(b) represents the Fermi surface (leftmost) and the constant energy contours at the same binding energies but taken at a temperature of 8 K which is below the AFM transition temperature. Upon careful inspection, one can clearly observe an  inner circular feature, absent in the paramagnetic phase, in the AFM phase which becomes more clear at 50 meV below the Fermi level . This feature grows in size as we go further below the  Fermi level asserting the hole-like nature of the bands and merges with the outer circular pocket up until 300 meV below the Fermi level. Going further towards higher binding energy, the merged feature completely vanishes and a new rectangular hole-like bulk feature begins to evolve in both paramagnetic and  AFM phases. However, the rectangular pocket in AFM phase is diminished as compared to the one in paramagnetic phase. This  indicates that the hole band is  pushed downwards in energy in the  AFM phase.

\noindent In Fig. 3, we present the dispersion maps along different momentum directions in both paramagnetic and AFM phases. Figure 3(a) shows the Fermi surface at 8 K (left) and 22 K (right) with white dashed lines representing the cut directions along which the dispersion maps are taken. Left panel in the Fig. 3(b) shows the band dispersion along the cut 1 (K-$\Gamma$-K) direction at 8 K, in which we can see two bands splitting very near the Fermi  level and merging going below the Fermi level all the way up to around 400 meV, consistent with the  feature observed in Fermi surface and energy contour plots in Fig. 2(b). The right panel in Fig. 3(b) shows the band dispersion along the K-$\Gamma$-K direction at 22 K, and it can be clearly seen that the hole-band is crossing the Fermi level and the extra feature near the Fermi level is absent, stressing the role of magnetism in the electronic structure near the Fermi level. In order to resolve these features clearly, we have presented the the second derivative plots of band dispersion along the K-$\Gamma$-K direction in Fig. 3(c).  Similarly, in Fig. 3(d), we present the dispersion maps along the cut 2 (M-$\Gamma$-M) direction where no extra feature is seen as in K-$\Gamma$-K direaction. 

\noindent In order to better visualize the changes due to magnetic ordering, we present the second derivative plot of constant energy contours at 50 meV below the Fermi level and momentum distribution curves (MDCs) along the K-$\Gamma$-K direction in Fig. 4. In  the second derivative plot of energy contour in Fig. 4(a) (AFM phase), we can clearly see a inner feature, which is absent in Fig. 4(b) (paramagnetic phase). This is the same feature which merges with the outer pocket at higher binding energies up to about 350 meV. Figure 4(c) shows the MDC plot along the K-$\Gamma$-K direction in the AFM phase. One can clearly see two peak features near the Fermi level which merge into one peak going further below the Fermi level. However, in the paramagnetic phase [Fig. 4(d)], we can only observe a single peak feature all the way up to Fermi level.
 
\noindent  Our study shows the signature of electronic structure change across the AFM transition temperature. In the paramagnetic phase, the two linearly dispersing hole bands cross the Fermi level, therefore the Dirac point is located above the Fermi surface. However, in the AFM phase, our ARPES data reveal the clear band splitting nature in the very vicinity of the Fermi level.   We suggest this observation as the possible signature of magnetic topological state on the lower part of the surface Dirac cone. One can bring the Dirac point at the Fermi level by tuning of chemical potential via the electrical gating or chemical doping. Our theoretical analysis shows that this compound is an axion insulator. Recent theoretical study suggests that this axion insulating state can co-exist with TCI state in \textit{in-plane}  and  HOTI phase in \textit{out-of-plane} configuration of the AFM phase \cite{dai}. Therefore, by tuning the magnetic moment configuration, our studied system could open up a new platform to study different types of topological states and their interplay with magnetism in this material.


\noindent \textit{Acknowledgement:} M.N. is supported by the Air Force Office of Scientific
Research under award number FA9550-17-1-0415 and
the National Science Foundation (NSF) CAREER award
DMR-1847962. The work at Northeastern University was supported by the
US Department of Energy (DOE), Office of Science, Basic Energy Sciences grant number DE-FG02-07ER46352, and benefited from Northeastern University's Advanced Scientific Computation Center (ASCC) and the NERSC supercomputing center through DOE grant number DE-AC02-05CH11231. D.K. was supported by the National Science Centre (Poland) under research Grant No. 2015/18/A/ST3/00057. We thank Sung-Kwan Mo for beamline assistance at the ALS, LBNL.

\end{document}